\documentclass[fleqn,usenatbib]{mnras}

\usepackage{newtxtext,newtxmath}

\usepackage{tikz}
\usetikzlibrary{positioning}

\usepackage[T1]{fontenc}

\DeclareRobustCommand{\VAN}[3]{#2}
\let\VANthebibliography\thebibliography
\def\thebibliography{\DeclareRobustCommand{\VAN}[3]{##3}\VANthebibliography}

\usepackage{graphicx}	
\usepackage{amsmath}	
\usepackage{footnote}
\usepackage[T1]{fontenc}
\usepackage{fonttable}

\newcommand{\kms}{\mbox{$\mathrm{km\,s}^{-1}$}}

\newcommand{\Msun}{\mbox{$\mathrm{M}_{\sun}$}}
\newcommand{\Rsun}{\mbox{$\mathrm{R}_{\sun}$}}
\newcommand{\Mwd}{\mbox{$M_\mathrm{wd}$}}
\newcommand{\Rwd}{\mbox{$R_\mathrm{wd}$}}
\newcommand{\Twd}{\mbox{$T_\mathrm{wd}$}}
\newcommand{\Porb}{\mbox{$P_\mathrm{orb}$}}
\newcommand{\Psh}{\mbox{$P_\mathrm{sh}$}}
\newcommand{\Psha}{\mbox{$P'_\mathrm{sh}$}}
\newcommand {\SDSSV}{SDSS\nobreakdash-V}
\newcommand{\allsdss}{SDSS~I~to~IV}

\usepackage{gensymb}

\title[V498 Hya, a new candidate for a period bouncer CV]{V498 Hya, a new candidate for a period bouncer Cataclysmic Variable}

\author[G. Tovmassian et al.]{Gagik Tovmassian$^{1}$,\thanks{E-mail: gag@astro.unam.mx}
Keith Inight$^{2}$,
Anna Francesca Pala$^{3,4}$,
Boris~T.~G\"ansicke$^{2}$,
Vedant Chandra$^{5}$, \newauthor
Matthew Green$^{6}$, 
Odette Toloza$^{7}$
and Matthias R. Schreiber$^{7}$
\\
$^{1}$Universidad Nacional Aut\'onoma de M\'exico, Instituto de Astronom\'{i}a, Aptdo Postal 106, Ensenada 22860, Baja California, M\'exico\\
$^{2}$Department of Physics, University of Warwick, Coventry, CV4 7AL, UK\\
$^{3}$European Southern Observatory, Karl Schwarzschild Stra{\ss}e 2, Garching, 85748, Germany\\
$^{4}$European Space Agency, European Space Astronomy Centre, Camino Bajo del Castillo s/n, 28692 Villanueva de la Cañada, Madrid, Spain\\
$^{5}$Center for Astrophysics | Harvard \& Smithsonian, 60 Garden Street, Cambridge, MA 02138, USA \\
$^{6}$School of Physics and Astronomy, Tel-Aviv University, Tel-Aviv 6997801, Israel \\
$^{7}$Departamento de F{\'i}sica, Universidad T{\'e}cnica Federico Santa Mar{\'i}a, Avenida Espa{\~n}a 1680, Valpara{\'i}so, Chile
}

\date{Accepted XXX. Received YYY; in original form ZZZ}

\pubyear{2015}

\begin{document}
\label{firstpage}
\pagerange{\pageref{firstpage}--\pageref{lastpage}}
\maketitle

\begin{abstract}
V498\,Hya (SDSS J084555.07+033929.2) was identified as a short-period cataclysmic variable (CV) by the  Catalina Real-Time Transient Survey (CRTS) in 2008. The superhump period was measured during the detected single superoutburst of V498\,Hya.  The quiescent spectrum subsequently taken by the \SDSSV\ Milky Way Mapper survey suggested that the CV donor may be a brown dwarf. We present time-resolved follow-up spectroscopy of V498\,Hya in quiescence, obtained with the GTC OSIRIS spectrograph, from which we derived the 86.053\,min spectroscopic period, systemic radial velocity, and the gravitational redshift of the \ion{Mg}{ii} line. We also modeled the spectral energy distribution to constrain the system parameters, including the $\ge 0.82 M_\odot$ mass of the white dwarf and the best-fit value $0.043\pm0.004\ \mathrm{M}_\odot$ of the donor star mass. This combination of parameters implies that V498\,Hya has evolved past the period minimum and is a relatively rare ``period bouncer''. 
\end{abstract}




\section{Introduction}

Cataclysmic Variables (CVs) are close binary stars containing a white dwarf accreting matter from a companion (hereafter the donor), which fills its Roche lobe. Several possible pathways can result in a binary star evolving into a CV \citep[see the recent review by][]{2023hxga.book..129B}. The main pathway starts with a common envelope phase, evolves into a semi-detached binary with a long period $>3$\,h, and then gradually drifts towards shorter periods. In this scenario, the donor is a main sequence star of  M or K ($\le0.8$\,\Msun) type, which does not significantly nuclearly evolve, i.e., its (core) composition does not change. However, its mass and radius will change a lot during the evolution of the CV. As the orbital period of a CV decreases to $\simeq80$\,min, the donor star's mass becomes insufficient to sustain core hydrogen burning, rendering it partially degenerate.
The donor star becomes electron-degenerate, and its radius increases with decreasing mass and 
hence, the orbital period also increases  \citep{paczynski_krzeminski_1979}. This turning point is called the period minimum. The exact value of the period minimum depends upon a number of factors, but it is approximately those mentioned above $80$\,min \citep{1999MNRAS.309.1034K,2011ApJS..194...28K}. 

Binaries that have evolved beyond the period minimum were dubbed period bouncers by \citet{1998PASP..110.1132P}. According to population models, a significant fraction of CVs should have become period bouncers with estimates ranging from 40--60 percent \citep{2015ApJ...809...80G}  to 70~per cent \citep{1993A&A...271..149K} or even  75~per cent \citep{2018MNRAS.478.5626B}. However, very few period bouncers have been identified. The reason for this shortfall is an open question - one possible explanation could be the impact on CV evolution of the white dwarf's magnetic field \citep{2023A&A...679L...8S}.  

\cite{2009MNRAS.397.2170G} demonstrated an anticipated spike of short-period CVs, arguing that they are intrinsically faint and, hence, were not discovered in large numbers before the Sloan Digital Sky Survey (SDSS). However, \citet{2009MNRAS.397.2170G} found little evidence of a period bouncer population in SDSS.
\citet{2011MNRAS.411.2695P} pioneered the search for period bouncers; he listed about two dozen candidates among the known CVs. Subsequently, \citet{2020MNRAS.494.3799P} surveyed all possible CVs within 150\,pc to minimize observational bias and found that the fraction of period bouncers was only 7--14~per cent within the limited volume.
Most recently, \citet{2023MNRAS.524.4867I} reported that only 2.6~per cent of the 507 CVs observed by \allsdss\, were period bouncers. 

\begin{figure}
	\includegraphics[width=\columnwidth]{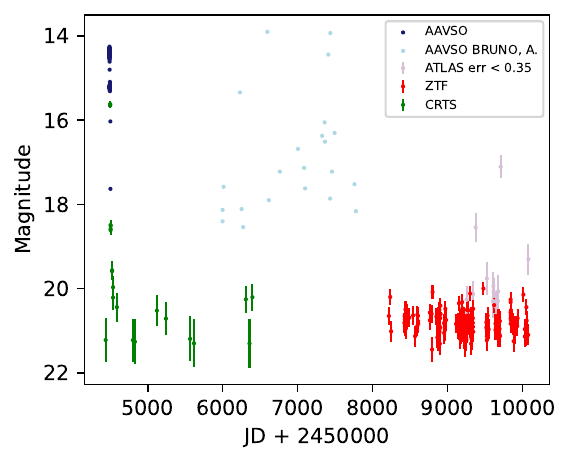}
    \caption{The light curve of V498\,Hya spanning over 5000 days from a variety of automatic sky surveys marked in the legend. One definite (multiple measurements) outburst was recorded throughout that time with a peak brightness V=15.62\,mag. The object demonstrates large ($\Delta m\simeq1.5$\,mag) variability in quiescence. Meanwhile, \citet{2008CBET.1631....1Y} reported a peak brightness of about $15.4$\,mag during the outburst and also reported the presence of superhumps (SHs), thereby qualifying this outburst as a superoutburst. The rare outbursts are consistent with V498\,Hya being a WZ\,Sge-type CV. Several other records indicating a possible outburst activity come from a single AAVSO observer (light blue points), and a close examination shows that they are not consistent with a regular outburst pattern. ATLAS data may also indicate frequent outbursts, but their errors are very large, and results are unreliable. Only measurements with errors $<0.29$ mag are included.  }
    \label{fig:lccrts}
\end{figure}

SDSS has now entered its fifth phase (\SDSSV,  \citealt{2017arXiv171103234K}), which will extend multi-object spectroscopy across the entire sky. This promises to produce the most complete census of CVs, and period bouncers in particular. A catalog of the 118 CVs observed in the first eight months of \SDSSV\  is reported by \citet{2023MNRAS.525.3597I} who describe the targeting strategy, the identification of new CVs, the spectral confirmation of candidate CVs, and new observations of previously known CVs. Based on well-known period bouncers with reliable astrometry and photometry, the authors also defined a space in the Hertzsprung-Russell diagram occupied by period bouncer systems and listed eight period bouncer candidate systems identified because their spectrum showed the signature of the white dwarf whilst not showing any sign of the donor. In addition to these eight in the catalog, there were other potential period bouncers with unreliable astrometry.  In particular, V498\,Hya was not considered because it did not satisfy the $G\le 20$ magnitude limit together with having a reliable parallax ($\Delta \varpi < 0.2 \times \varpi$) defined by \citet{2023MNRAS.525.3597I}.

V498\,Hya was discovered prior to \SDSSV\, by \citet{2008CBET.1631....1Y} due to the occurrence of a dwarf nova superoutburst (Fig.\,\ref{fig:lccrts}) on 2008 January 19 ($\mathrm{JD}=2\,454\,487$). The Catalina Real-Time Transient Survey (CRTS, \citealt{2009ApJ...696..870D}) also captured the superoutburst.  The CRTS recorded no additional outbursts, and there is no evidence of outbursts in the ASAS-SN database \citep{2014ApJ...788...48S,2017MNRAS.467.1024M}.

However, the AAVSO reported sporadic observations in a time period $\mathrm{JD\,2\,456\,000\ and\ 2\,458\,000}$\ in TG filter\footnote{Green Filter (or Tri-color green). This is commonly the "green-channel" in a CCD camera or Astroimaging G filter. These observations use V-band comp star magnitudes.} show the object generally brighter than it is in the quiescence and two or three outbursts-like events. \citet{2021MNRAS.502.5668V} even determined the superoutburst cycle lengths of V498\,Hya based on these observations. 

The light curve of V498\,Hya, spanning several years (Fig.\,\ref{fig:lccrts}), shows a reported variability of 
$\sim1.5$\,mag during quiescence based on AAVSO data. However, this variability is inconsistent with higher-quality observations from ZTF, which shows a much narrower spread of flux measurements, with a standard deviation of approximately 12\% in the 
$g$- and $r$-bands.

We attribute the reported variability to poor data quality, including uncertain or single-observer AAVSO measurements. The database flagged these observations as "uncertain," and there is a lack of follow-up monitoring typically associated with outbursts. As a result, we exclude the AAVSO data from further analysis and confirm that the quiescence variability is minimal and consistent with expectations for a faint WZ Sge-type CV.

 ATLAS data \citep{Heinze_2018} also reports a multitude of bright states with very large errors. Filtering them down to mag.err$<0.35$ leaves only a few bright points, and with a stronger criteria mag.err$<0.29$ no points at all. Given a relatively low spatial resolution and small sensitivity, we also discard these as outburst detections.

Analysis of the light curve of V498\,Hya during the initial stages of the superoutburst revealed superhumps (SHs) with a mean period  $P_\mathrm {sh}=0.06036\pm0.00002$\,d  \citep{2009PASJ...61S.395K}. Based on that data, \citet{2012PASJ...64...63K} classified the object as a probable SU\,UMa type. \citet{2023MNRAS.525.3597I} used over a decade of additional photometry from ZTF and CRTS to conclude that V498\,Hya most probably belongs to the WZ\,Sge type, a sub-class of the SU\,UMa type \citep{2015PASJ...67..108K}. Whereas SU\,UMa type stars generally have frequent superoutbursts each followed by a series of regular dwarf nova outbursts,  WZ\,Sge have infrequent and unusually large superoutbursts ($\Delta m \simeq 7$\,mag). 
WZ\,Sge CVs also typically have a low mass ratio (donor mass divided by white dwarf mass) and low mass-transfer rate, which are both indicative of an old CV with a low mass donor \citep{1979AJ.....84..804P,2010MNRAS.402.1816C}. These properties are consistent with being period bouncers, but not all WZ\,Sge stars are necessarily period bouncers.

\section{Observations}
\subsection{OSIRIS}
Spectroscopic observations were performed with the 10.4\,m Gran Telescopio Canarias (GTC) at the Observatorio del Roque de los Muchachos
(La Palma, Spain). V498\,Hya was observed in service mode using two 1.5-hour observing blocks on two consecutive nights in 2022 November 2 and 3. The spectra were taken with the Optical System for Imaging and low-intermediate-resolution Integrated Spectroscopy (OSIRIS) spectrograph \citep{2003SPIE.4841.1739C}, which consists of a mosaic of two Marconi CCDs, each with $2048\times4096$ pixels. Each pixel has a physical size of 15\,$\mu$m. We used $2\times2$ binning  for our observations, giving a plate scale of 0.254\,arcsec. OSIRIS was used in long-slit mode, centering the objects in CCD2. 
We used the R2000B volume-phased holographic grating, providing $\lambda 3950 - 5700$\,\AA\ coverage at $R=2165$ resolution.  Twenty-two spectra of the object were obtained (eleven each night) with 480\,s exposure times in dark spectroscopic conditions with 0.8\,arcsec seeing; the slit width was set to 1.0\,arcsec.

\begin{figure}
\includegraphics[width=\columnwidth, bb=0 0 270 220, clip=]{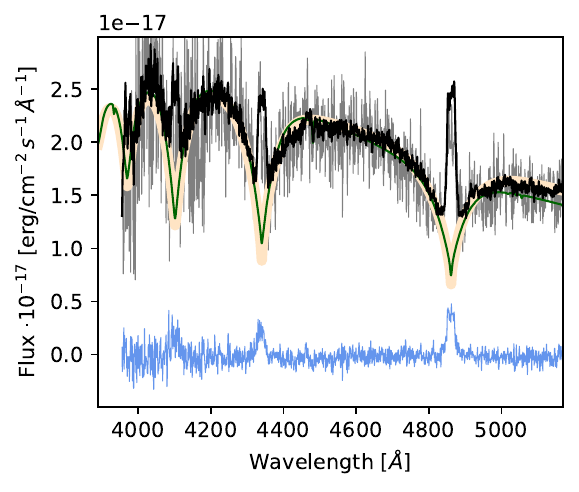}
\caption{OSIRIS spectra of V498\,Hya. The black line shows the average spectrum of the 22 individual exposures. Two white dwarf models are over-plotted. The initial model with $T_{\mathrm{wd}}=13\,000$\,K and $\log g=9.0$ by \citet{2010MmSAI..81..921K} is shown in orange whilst the final model with $T_{\mathrm{wd}}=18\,100$\,K 
is shown in green.  The contribution of the continuum from the disk becomes noticeable with $\lambda>5000$\,\AA, and is modeled in Section\,\ref{sec:fit}.
A grey line shows an example of a single exposure, and the residual emission-line spectrum after subtracting the model white dwarf spectrum from the average is shown in blue.}
    \label{fig:gtcsp}
\end{figure}

The OSIRIS spectra were reduced using \textsc{IRAF} v2.16 \citep{1986SPIE..627..733T,1993ASPC...52..173T}. We followed the standard data reduction procedure for long-slit spectroscopy. The spectra were wavelength-calibrated with Hg-Ar, Ne, and Xe lamps. Barycentric corrections were applied as part of the standard data reduction procedure. The flux calibration was achieved by observations of the standard star G191-B2B. Fig.\,\ref{fig:gtcsp} shows the averaged spectrum composed of all twenty-two individual exposures.

\begin{figure*}
\setlength{\unitlength}{1mm}
\resizebox{12.5cm}{!}{
\begin{picture}(100,80)
\put (45,-15){\includegraphics[width=72.5mm, bb=0 0 564 690, clip=]{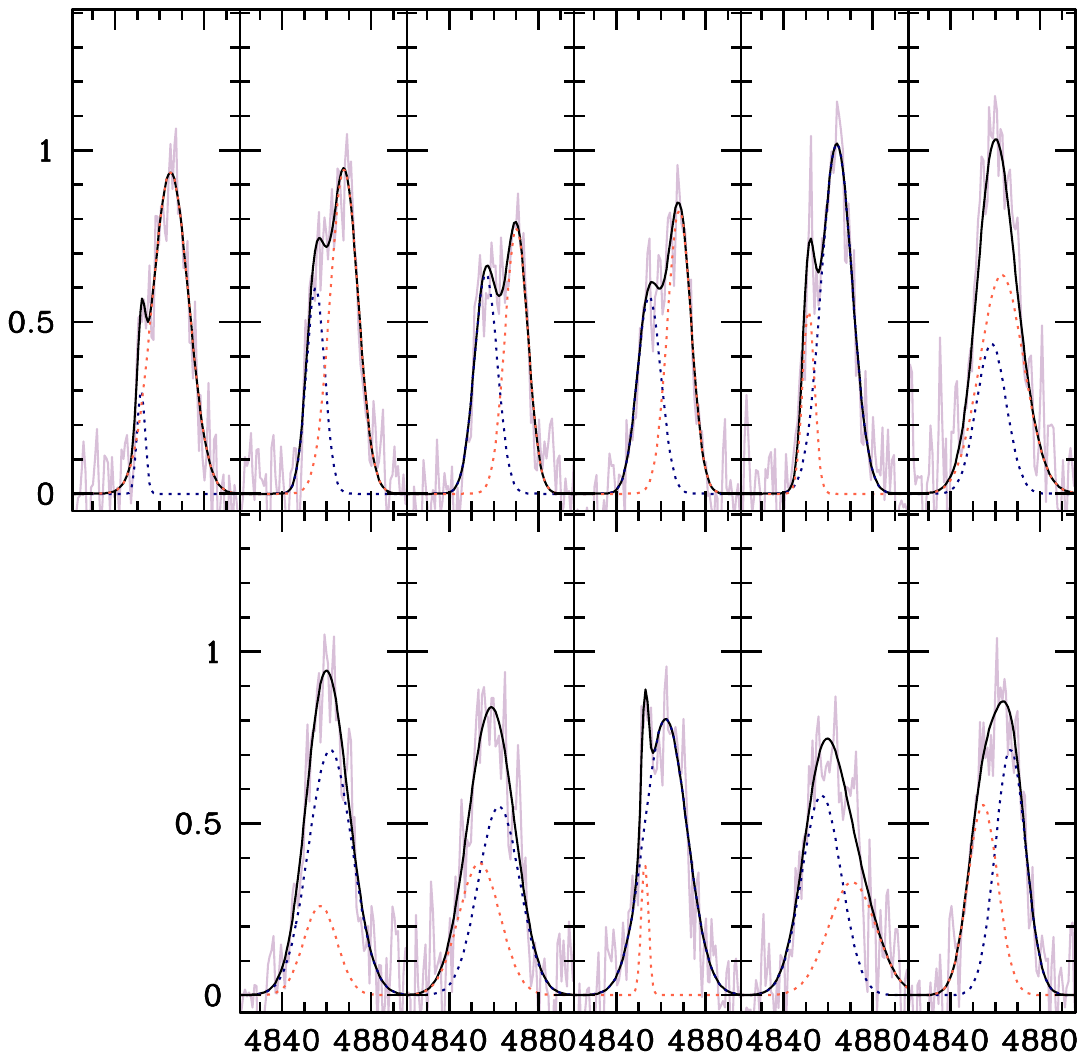}}
\put(-19,30){\rotatebox{90,0}{\ {\fontfamily{DejaVuSans-TLF}\selectfont Relative Flux}}}
\put(46,2){\makebox(0,0)[l]{\fontfamily{DejaVuSans-TLF}\selectfont Wavelength [\AA]}}
\put (-19,-15){\includegraphics[width=72.5mm, bb=0 0 564 690, clip=]{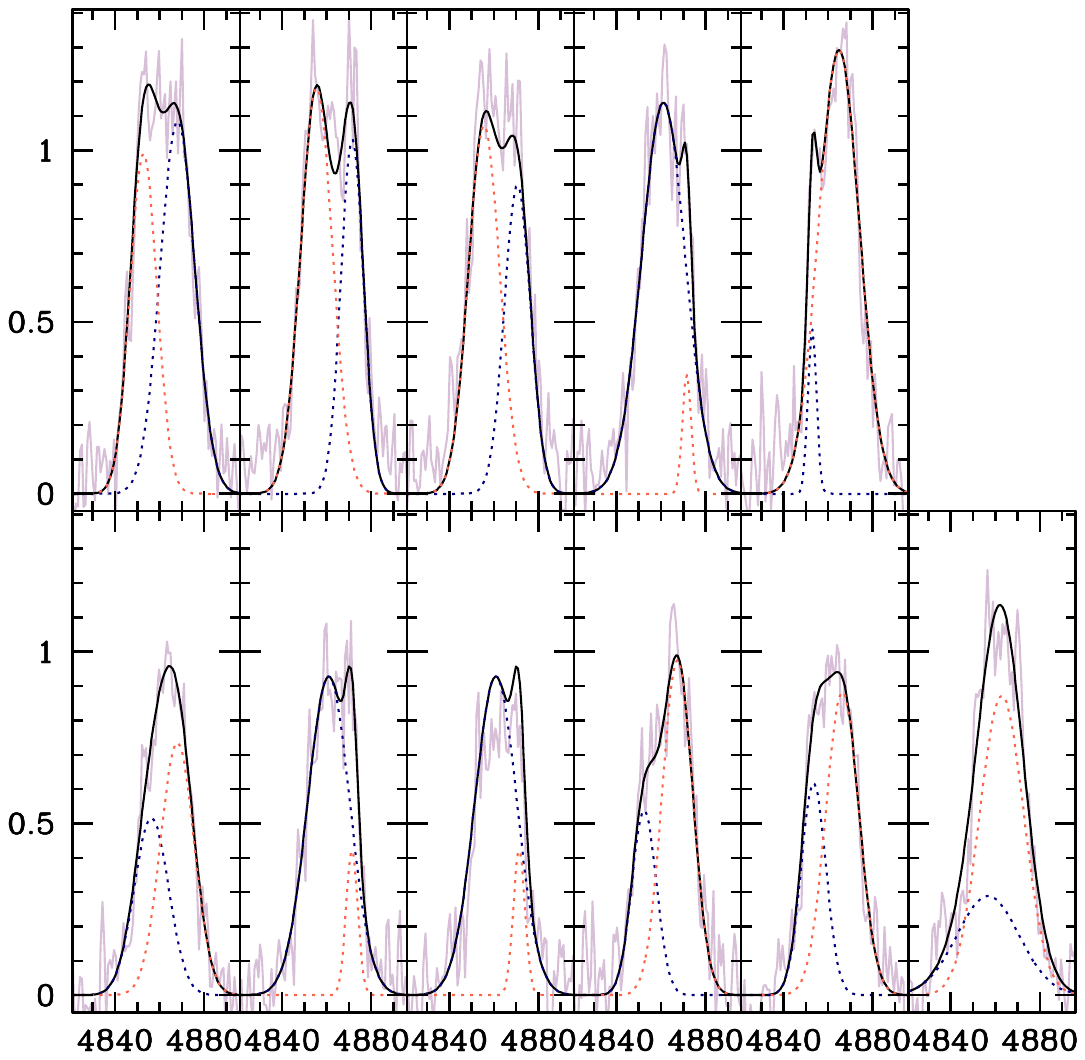}}
\end{picture}}
 \caption{The 22 OSIRIS spectra of V489\,Hya after subtracting a white dwarf model are shown in light violet. Each spectrum was fitted using two Gaussians (overplotted in red and blue), and the sum of the two Gaussians is shown in black.} \label{fig:deblending}
\end{figure*}
\subsection{SDSS}
We also used the object's \SDSSV\ DR18 spectrum for this study. It was obtained by the SDSS BOSS spectrograph \citep{2013AJ....146...32S,2013AJ....145...10D} and covering the range  $3600-10\,400$\,\AA. The BOSS data was processed with the \SDSSV\ v6\_0\_4  pipeline \citep{2023ApJS..267...44A}, which performed sky subtraction together with the flux and wavelength calibration of the spectrum.

We used the VizieR Catalogue to access photometric spectral energy distribution (SED) data 
for this object \citep{10.26093/cds/vizier}.
In addition to the VizieR data, we use an important for this study IR detection of V498\,Hya by the UKIDSS Large Area Survey (LAS) \citep{2018MNRAS.473.5113D}\footnote{The UKIDSS project is defined in \citet{2007MNRAS.379.1599L}. UKIDSS uses the UKIRT Wide Field Camera \citep[WFCAM;][]{2007A&A...467..777C} and a photometric system described in \citet{2006MNRAS.367..454H}. The pipeline processing and science archive are described in  \citet{2008MNRAS.384..637H}} The measurements are available in Y ($20.64\pm0.24$) and J bands ($19.77\pm0.2$), whether the object was not visible in H \& K bands.

\subsection{\textit{Swift}}\label{subsec:swift}
\textit{Swift} UVOT \citep{2005SSRv..120...95R} observations were taken to expand our spectral range into the UV, where the white dwarf dominates. Observations were taken on 2023, September 9 and 10, using the UVW1, UVM2 \& UVW2 filters (ObsID 16220001) with total exposure times of $2266.5$\,s, $2412.0$\,s, and $2412.0$\,s respectively. The data were locally reprocessed by the UK {\it Swift} Science Data Centre using \textsc{HEASoft v6.32}. The source was not detected in any of the filters, and so the flux limits were calculated using the UVOT Tool for Simulating Observations with Point Sources\footnote{
\url{(https://www.mssl.ucl.ac.uk/www_astro/uvot/uvot_observing/uvot_tool.html)}} assuming a white dwarf source. Corresponding minimum visual magnitudes for a $5\sigma$ detection are $UVW1=22.39$\,mag, $UVM2=22.34$\,mag, and $UVW2=22.84$\,mag. For a $10\,000$\,K blackbody they translate into $1.365 \times 10^{-16}$, $1.822 \times 10^{-16}$ and $2.189 \times 10^{-16}$ erg\,cm$^{-2}$\,s$^{-1}$\,\AA$^{-1}$ respectively \citep{2016yCat..51520102B}.

\section{Analysis}
To determine whether V498\,Hya is a period bouncer, we need to assess its system parameters and, in particular, the mass of the donor star. Since the donor is not detectable in the spectrum, we have to use an indirect method for estimating the donor mass.

The first step is to use a relation between the SH period excess (the difference between the SH period and the actual orbital period - see Section \ref{sec:shumpexcess}) and the ratio of donor mass to white dwarf mass ($q = M_\mathrm{donor}/M_\mathrm{wd}$) (see \citealt{1984A&A...132..187S,1984ApJS...54..443P}\, and also \citealt{2022arXiv220102945K}\, for the latest comprehensive review and additional references). This requires an accurate estimate for the orbital period (Section\,\ref{sec:period}). 
The second step is to derive the white dwarf mass by estimating the gravitational redshift (Section\,\ref{subsubsec:gr}), which in turn requires an estimate for the systemic velocity (Section\,\ref{sec:systemic}). The third step is to fit a model of the CV to the spectral energy distribution (Section\,\ref{sec:fit}).

\subsection{Superhump period excess}\label{subsec:q}
\label{sec:shumpexcess}

The emergence of humps in the light curve during the early stages of a superoutburst of WZ Sge-type CVs was discovered by \citet{1981ApJ...248.1067P,1996PASP..108..748P}. These early humps have either been called outburst orbital humps by \citet{1996PASP..108..748P} or early SH \citep{1996PASJ...48L..21K}.
\citet{2022arXiv220102945K} shows that the relation between the mass ratio ($q$) and the period excess ($\epsilon={\Psh/\Porb} - 1$) depends on when the SH period was measured. \citet{2022arXiv220102945K} defines Stage\,A as lasting for the initial $\simeq20$\, orbital cycles followed by Stage\,B. Note that the mass ratio determination using the period excess during the stage\,B is not as reliable because of the pressure effect influencing the mass distribution of the disk, and it is difficult to formulate \citep{2001MNRAS.325..761M,2006MNRAS.371..235P}. 

\citet{2009PASJ...61S.395K} reported the SH period of V498\,Hya to be $\Psh=0.06036\pm0.00002$\,d (=86.9184\,min) It was determined between the 66th and 167th orbital periods after the superoutburst. The SHs were, therefore, most likely observed during the stage\,B. 

We requested and analyzed the data provided by Taichi Kato to independently assess the SH period of V498 Hya. Our analysis confirmed the presence of two significant peaks in the power spectrum: one corresponding to 
$\Psh=0.06036$\,d, as reported by \citet{2009PASJ...61S.395K}, and another at $\Psha=0.06230$\,d or 89.716\,min. The folded light curves suggest that the first period produces a slightly smoother profile, which aligns more closely with the S-wave period derived from hotspot analysis. This consistency supports its selection as the preferred solution, although the alternative cannot be conclusively excluded.

Given the broad peaks and inherent scatter in the original dataset, we consider both solutions cautiously in our discussion.

\subsection{The Orbital Period}
\label{sec:systemic}

The GTC spectra of V498\,Hya (Fig.\,\ref{fig:gtcsp})  show deep, broad hydrogen absorption features from the white dwarf superposed by Balmer emission lines.
Balmer emission lines, formed in the accretion disk, are commonly used to measure CVs' radial and systemic velocities. 
The simplest but rough method is to fit the emission lines with a Gaussian profile and use RVs to determine the orbital period. A discreet Fourier transformation (DFT) period search on such measurements results in a peak at $f=17.33$\,c/d, which is probably a one-day alias, since from the SH period, we expect the orbital period to be in 16’s c/d. We can fit the measured RV by a $-1$ day alias and then improve the fit by varying the frequency using the Period04 software \citep{2005CoAst.146...53L}. This results in a best fit with $f=16.35$\,c/d and 140 km/s semi-amplitude of RVs. 
We obtained a similar value by cross-correlating a template made by an averaged combination of evenly-phased twenty-two spectra with each spectrum.

The widely deployed double-Gaussian method \citep{1980ApJ...238..946S} measures the radial velocity in the wings of the line reflecting the orbital motion of the internal ring of the disk closest to the white dwarf. 
This method needs well-defined wings in the lines, which usually are distinct from the underlying flat (power law) continuum in long-period systems. This is not the case with V498\,Hya. The contribution from the white dwarf is very large, so the absorption significantly affects the wings of emission lines.

We averaged all available GTC spectra, improving the signal-to-noise ratio (S/N) to 25 around H$\beta$. We then assumed a relatively cool white dwarf, as typical for this period range \citep{2017MNRAS.466.2855P}, $T_{\mathrm{wd}}\lesssim15\,000$\,K,  and found that a $T_{\mathrm{wd}}=13\,000$\,K $\log g=9.0$ \citep{2010MmSAI..81..921K} white dwarf model provides a good match to the average spectrum (orange line in Fig.\,\ref{fig:gtcsp}). This model was subtracted from the individual spectra we observed\footnote{We subsequently determined $T_{\mathrm{wd}}=18\,000$\,K (section\,\ref{sec:fit}) but as Fig.\,\ref{fig:gtcsp} shows the difference in the subtracted spectra will not be significant.}.   This resulted in a set of spectra with emission lines towering over the flat continuum (for example, the blue line in Fig.\,\ref{fig:gtcsp}).  

After performing the double-Gaussian procedure on the white dwarf subtracted spectra, we obtained a very low semi-amplitude of RVs, measuring about 8\,\kms\, which was inadequate\footnote{The resolution in velocity for a spectral $R=2165$ R2000B grism at $\lambda4755 \AA$ is approximately 140 km/s. Since we have 44 measurements per orbital cycle, the minimal measurable semi-amplitude of RV  would be $140/\sqrt(44)\approx21$\,km/s. } for determining the orbital period. Curiously, low-amplitude RV variations can be nicely folded with a period corresponding to the $f=16.31$\,c/d frequency.

We can not claim that our obtained RV semi-amplitude and, hence, the inferred period are correct\footnote{SDSS J105754.25+275947.5 \citep{2023MNRAS.526.5110E} is an excellent example of successful measurement of a similar but higher inclination object observed with the same telescope/instrument but with a slightly higher spectral resolution.}. However, it allows us to deduce the value of the systemic velocity \mbox{$\gamma=-13\pm1.7$\,\kms}\ from the narrow spread of points. 

\subsection{Bi-modal solutions}
\label{sec:period} 

\begin{figure*}
\includegraphics[width=\textwidth, bb=60 70 680 320, clip=]{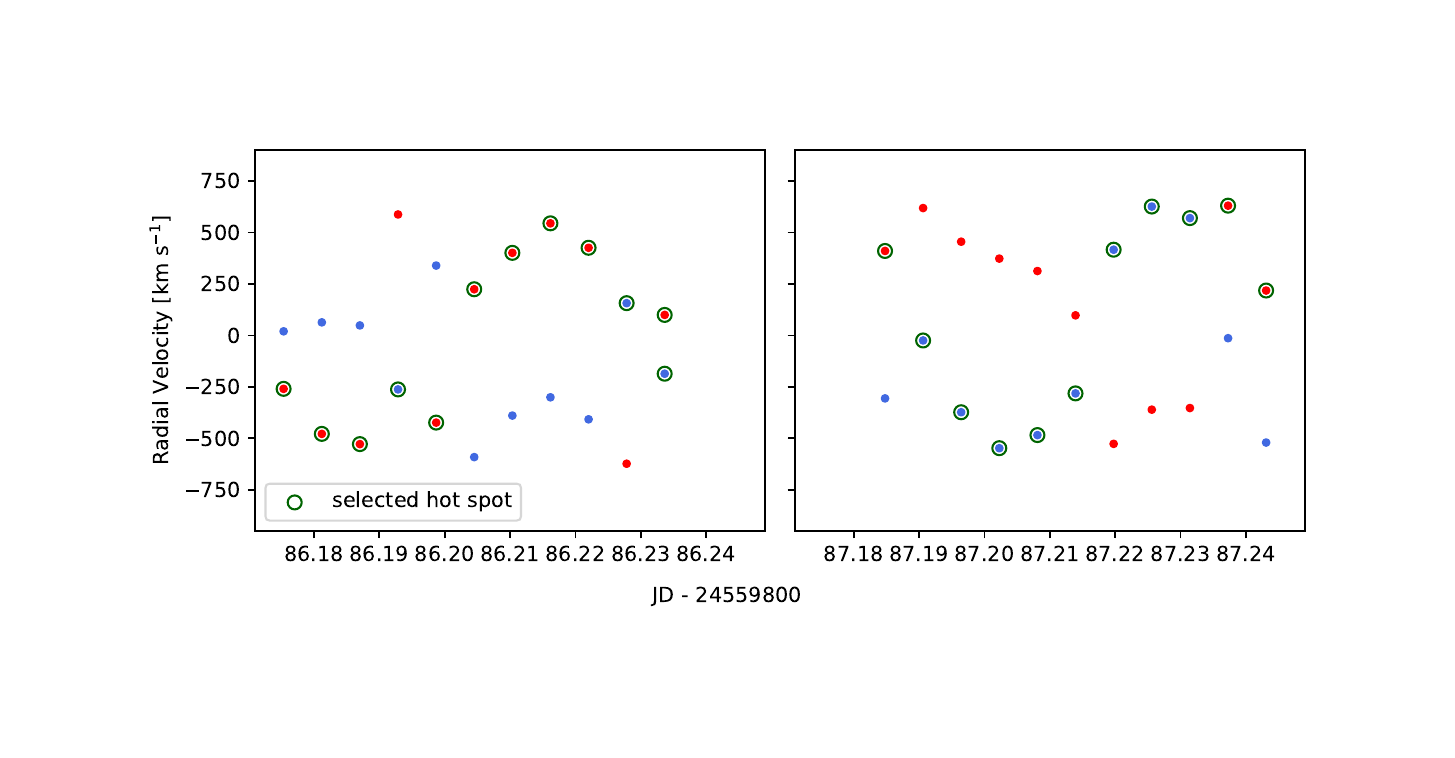}
\caption{The RV measurements of the two components from the de-blending procedure of the H$\beta$ emission line are plotted (red and blue dots) in two panels, each corresponding to one night. Additionally, we marked the points we identified as pertaining to the hot spot with green circles. The choice is somewhat arbitrary, but we were guided by choosing a set of points forming a "perfect" sinusoidal wave. } 
\label{fig:rvhs}
\end{figure*}

The H$\beta$ emission is the only well-exposed line in our GTC spectra. Unlike the familiar double-peaked shape commonly observed in high-inclination CVs, it has a more complex profile, as indicated by the light violet in Fig.\,\ref{fig:deblending}). There is a component, however, that moves back and forth at the tip of the line. 
We implemented a novel approach, making use of the fact that emission lines in CVs often feature an S-wave corresponding to the hot spot (or bright spot) where the mass-transfer stream collides with the accretion disk. The remainder of the emission comes from the rest of the disk \citep{1986MNRAS.218..761H}.  We used the {\sl deblend} option in the \textsc{IRAF} {\sl splot} procedure following the methodology used in  \citet{2018ApJ...869...22T,2021MNRAS.503.1431H} and elsewhere. 
The process fits two Gaussians to each of the H$\beta$ profiles based on the assumption that one Gaussian is modeling the hot spot and the other Gaussian is modeling the residual (see Fig.\,\ref{fig:deblending}). The wavelength of the centre of each Gaussian is converted to a radial velocity (RV). The set of RV measurements (two per spectrum) is shown in  Fig.\,\ref{fig:rvhs}. Since we know that the hot spot contribution is a significant part of the emission line profiles of thin disks, we selected a set of RV measurements of relatively even widths and intensities to form a sinusoidal pattern. 
We marked them with green circles, assuming they corresponded to the hot spot. 
The procedure is somewhat arbitrary. In particular, it is possible to include either or both of the last RVs on the first night JD\,86.235, causing a slight ($\Delta f \simeq 0.02$) variation in the calculated period. Henceforth, our analysis assumes both points are included in the fit.

\begin{figure}
	\includegraphics[width=\columnwidth, bb=0 5 280 240, clip=]{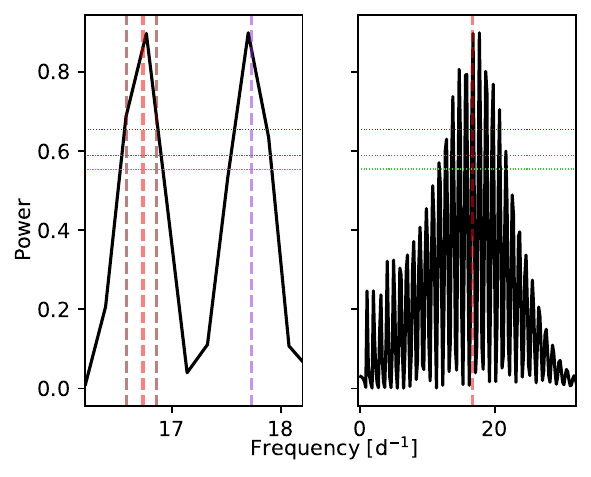}
    \caption{Lomb-Scargle periodogram of selected RV measurements of the hot spot. In the right panel, a wide range of frequencies is shown. The left panel focuses upon the two tallest peaks in the power. Horizontal dotted lines indicate false alarm probability limits of $0.1$, $0.05$ and $0.01$. Brown vertical dashed lines mark the range of frequencies with false-alarm probabilities less than  0.01. A red dashed line marks our preferred orbital frequency. The purple vertical dashed line indicates a $1\,\mathrm{d}^{-1}$\ alias.}
    \label{fig:lsrvhs}
\end{figure}

\begin{figure}
	\includegraphics[width=\columnwidth, bb=0 5 300 220, clip=]{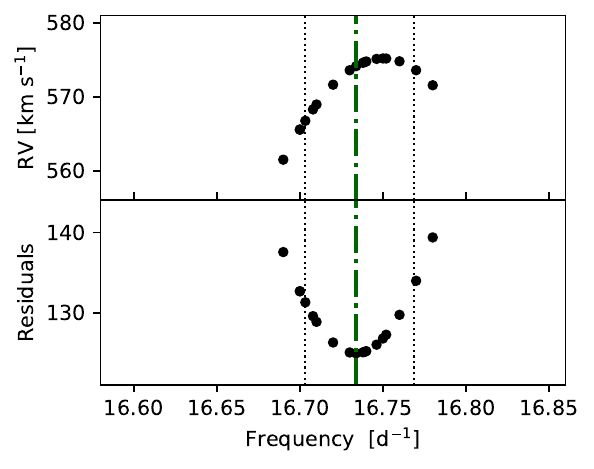}
    \caption{The variation of the RV amplitude and the residual from the fit as a function of the chosen orbital frequency. An insignificant difference exists between the minimum residuals and the maximum RV amplitude. Dotted vertical lines mark a narrow range of best-fit frequencies. A vertical dashed-dotted line marks our chosen solution with the smallest residuals.}
    \label{fig:frrange}
\end{figure}

\begin{figure}
	\includegraphics[width=\columnwidth, bb=0 5 300 220, clip=]{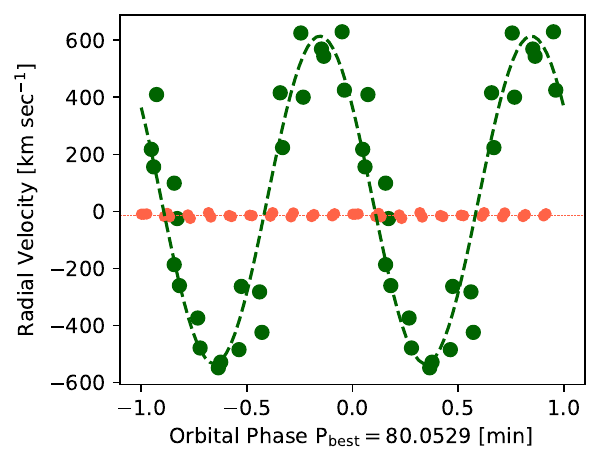}
    \caption{The RV measurements (green, large points) of the S-wave component of the H$\beta$ emission line corresponding to the "hot spot." They were fitted by a sinusoid with a semi-amplitude of 574\,\kms. Red points are velocities obtained by the double-Gaussian method; the horizontal dashed line is the average velocity -13\,\kms, representing the systemic velocity of the system.}
    \label{fig:frvhs}
\end{figure}

The power spectrum of the hot spot RVs is displayed in Fig.\,\ref{fig:lsrvhs}. A vertical red dashed line marks the maximum power. Unfortunately the peak is wide with $16.58 < f < 16.86\,\mathrm{d}^{-1}$ for frequencies with false alarm probabilities $<0.01$.  To improve the solution, we fitted a set of sinusoids to the series with the expectation that the optimum period would coincide with the maximum amplitude and minimum residuals. The results are plotted in Fig.\,\ref{fig:frrange}.
They show that the preferred frequency lies in the range $16.703<f<16.769\,\mathrm{d}^{-1}$. The value with the smallest residuals is $16.73389\ \mathrm{d}^{-1}$ (corresponding to the $P_{\mathrm{best}}=86.053 \pm 0.159$\,min; see Fig.\,\ref{fig:frrange} and Table\,\ref{tab:qurange}), which we adopt as the most probable orbital period of the system.
The corresponding RV measurements folded with the preferred frequency and their fit are shown in Fig.\,\ref{fig:frvhs}.

Using white dwarf-subtracted spectra, we calculated a Doppler tomogram \citep{1988MNRAS.235..269M,1998astro.ph..6141S}. 
The Doppler map presented in Fig.\,\ref{fig:dopmap} has the hallmark of a typical dwarf nova \citep{1998MNRAS.299..768S} with a marked ring corresponding to the accretion disk and a compact, bright spot, as routinely observed in other similar systems \citep{2018MNRAS.481.2523P,2021ApJ...918...58A,2023MNRAS.523.6114N,2023MNRAS.526.5110E}, confirming our assumption of the presence of a hot spot component in line profiles.

\begin{figure}
\setlength{\unitlength}{1mm}
\resizebox{9.6cm}{!}{
\begin{picture}(100,80)
\put (3,3){\includegraphics[width=\columnwidth, bb=26 40 620 578, clip=]{Figures/V498_Figure08.pdf}}
\put(0,30){\rotatebox{90,0}{\ {\fontfamily{DejaVuSans-TLF}\selectfont Vy [km s$^{-1}$]}}}
\put(40,0){\makebox(0,0)[l]{\fontfamily{DejaVuSans-TLF}\selectfont Vx [km s$^{-1}$]}}
\end{picture}}
\caption{Doppler map of the H$\beta$ line demonstrates the accretion disk's presence with a bright spot (bright red) at the impact point of the mass-transfer stream with the disc. }
\label{fig:dopmap}
\end{figure}

In conclusion, we obtained two sets of SH and orbital periods. Given what we know about short-period CVs, we must find the right combinations that make sense. 
In Table\,\ref{tab:qurange}, we present both sets of mass ratios $q$ deduced from the possible SH period excess, assuming the range of possible orbital periods. We are guided by the expectation that, at early stages, the normal SH period is always longer than the orbital period. In the bottom part of the table, we have a less favorable alternative SH period coupled with one-Gaussian RV measurements, which can not be precise taking into account the complex profiles of the emission lines, and two-Gaussian measurements, which provide dubious RVs, since the spectral resolution of observations is not good enough to determine velocities that as low as were obtained. Our preference is for the solutions provided by the measurements of the S-wave caused by the hot spot, hence the smallest mass $q<0.73$\ ratios.

\begin{table}
\renewcommand{\arraystretch}{1.3} 
\caption{Mass ratio ($q$) derived from $\epsilon$ using the relation from \citet{2022arXiv220102945K} \\ }
\centering
\begin{tabular}{llcccc}
\hline \hline
    Period & Frequency ($\mathrm{d}^{-1}$) & $\Porb$ (min)  & $\epsilon^1$  &$q$ \\  
\hline
    \Psh & 16.56726 & 86.92 &  &  & \\
    \Porb$^2$ min    & 16.76980 &  85.87 & 0.012 & 0.073\\
    \Porb$^2$ the best     & 16.73389 &  86.05 & 0.010 & 0.066 \\
    \Porb$^2$ max     & 16.70330  & 86.21 &   0.008 & 0.060\\
\hline
    \Psha & 16.0507 & 89.716 &  &  & \\
    1 Gaussian    & 16.35 &  88.07 & 0.018 & 0.094 \\
    2 Gaussian     & 16.3088 &  88.296 & 0.016 & 0.08 \\
\hline \hline
\multicolumn{4}{l}{
\small $^1\,\epsilon = \Psh/\Porb - 1$ } \\
\small $^2$\  S-wave \\
\end{tabular}    
\label{tab:qurange}
\end{table}

\subsection{Gravitational redshift}\label{subsubsec:gr}
Photons leaving the extreme gravitational field of the white dwarf are redshifted, and we seek to quantify this in terms of radial velocity. In addition, the gravitational redshift velocity is a function of the mass and radius \citep{1967AJ.....72Q.301G} of the white dwarf, allowing a measurement of the white dwarf mass. 

\begin{figure}
	\includegraphics[width=\columnwidth]{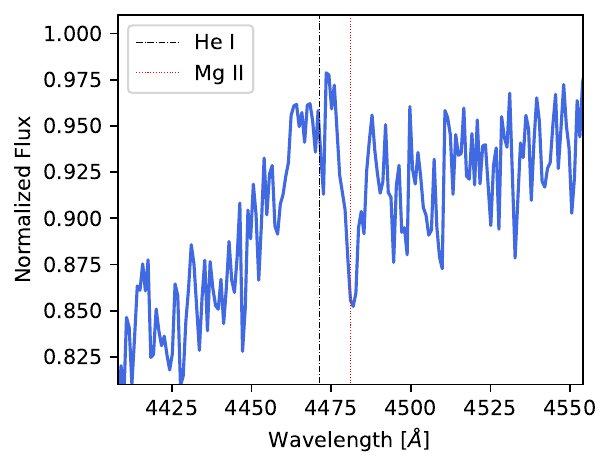}
    \caption{\ion{Mg}{ii} absorption line in the integrated spectrum of V498\,Hya. Also marked are the rest wavelength of \ion{He}{i} emission line and \ion{Mg}{ii}~4481\,\AA\ by dashed-dotted and dotted vertical lines, respectively. }
    \label{fig:mg2}
\end{figure}

In the integrated spectrum of V498\,Hya, it is possible to detect the presence of the \ion{Mg}{ii}~4481\,\AA\ absorption line (upper panel of Fig.\,\ref{fig:mg2}).
The \ion{Mg}{ii} absorption line originates in the photosphere of the accreting white dwarf and hence is subject to gravitational redshift.

The systemic velocity ($\gamma$) and gravitational redshift ($\nu_{\mathrm{grav}}$) are critical parameters. Still, both are subject to substantial uncertainties due to the limited resolution of our observations and the complexity of the \ion{Mg}{ii}\ $\lambda 4481$\,\AA\ absorption line profile. Using the central wavelength of the \ion{Mg}{ii}\ line, we estimate
$\nu_{\mathrm{obs}}=49.5$\,\kms, with a formal measurement error of approximately 
$\pm13$\,\kms. Subtracting the systemic velocity of $\gamma=-13$\,\kms,
we derive $\nu_{\mathrm{grav}}=62.5$\,\kms. The precision of these values is limited by the spectral resolution and the broad width of the \ion{Mg}{ii}, which exceeds the instrumental profile by a factor of two. Consequently, while these estimates provide a basis for calculating the white dwarf mass (M$_{\mathrm{WD}}=0.89\pm0.12$\,\Msun), they are best regarded as indicative. This approach aligns with our broader methodology, which emphasizes conservative error margins to avoid over-interpretation of uncertain measurements.

\begin{figure}
\includegraphics[width=\columnwidth]{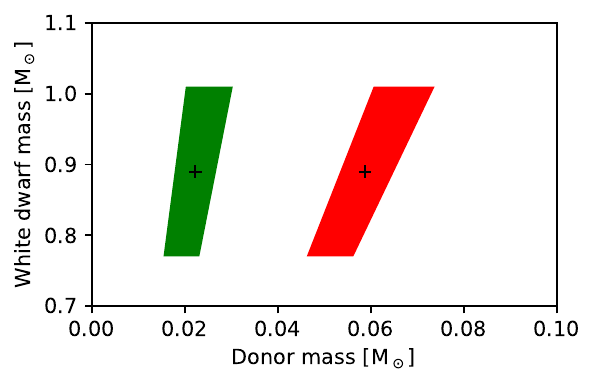}
    \caption{Plot of potential solutions (taking account of uncertainties) for the white dwarf and donor masses where $q$ is derived from the period excess and the white dwarf mass from the gravitational redshift. The green area assumes Stage\,A SHs, and the red area assumes Stage\,B.  }
    \label{fig:massvsmass}
\end{figure}

The possible donor star masses range is shown in Fig.\,\ref{fig:massvsmass}. With the orbital period of 0.0598\,d ($86.05\pm0.18$\,min), the upper limit of the donor star mass still places V498\,Hya among the period bouncers rather than the systems still approaching period minimum \citep[see for example Fig.\,1 in ][]{2022arXiv220102945K}.

\subsection{Spectral Energy Distribution and overall fit}
\label{sec:fit}

\begin{figure*}
	\includegraphics[width=\textwidth]{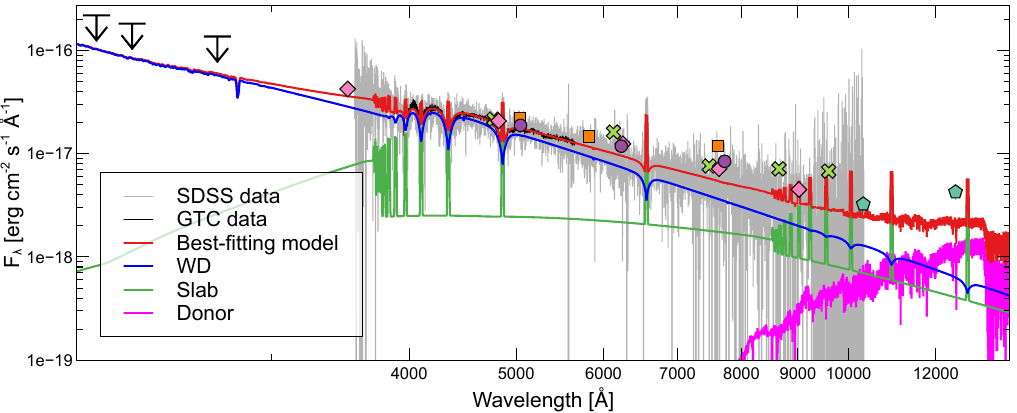}
    \caption{SDSS (grey) and  GTC (black) spectra of V498\,Hya along with the best–fit model (red, see Table~\ref{tab:fit_params} for the relevant parameters), which is the sum of a white dwarf (blue, $T_{\mathrm{wd}} = 18\,100$\,K), an isothermal and isobaric pure–hydrogen slab (green) and a late-type star (magenta). VizieR photometric data points are plotted (see Table~\protect\ref{tab:photometry}, circles for \textit{Gaia} DR2, squares for \textit{Gaia} EDR3, diamonds for SDSS-DR16, crosses for Pan-STARRS and pentagons for UKIDSS). 
    The upper limits of UV data from {\sl Swift}-UVOT are shown with the black arrows (see Section~\ref{subsec:swift}.)}
    \label{fig:sed}
\end{figure*}

\begin{table}
   \centering
   \caption{Vizier photometry of V498\,Hya. The \textit{Swift} upper limits are flagged with downward arrows. The UKIDSS data points were included in the spectroscopic fit (Section~\ref{sec:fit}) to complement the wavelength range covered by the spectroscopic observations.}
   \begin{tabular}{lccc} 
      \hline \hline
	Survey & Filter & Wavelength & Flux                  \\
               &        & $\AA$    & [$10^{-17}$ erg cm$^{-2}$ s$^{-1}$ $\AA^{-1}$] \\
      \hline
        \textit{Swift}/UVOT & \textit{UVW2} & 2083.95 & 21.9$\downarrow$ \\
        \textit{Swift}/UVOT & \textit{UVM2} & 2245.03 & 18.2$\downarrow$ \\
        \textit{Swift}/UVOT & \textit{UVW1} & 2681.67 & 13.6$\downarrow$ \\[0.1cm]
        \textit{Gaia} DR2   & $G_\mathrm{BP}$ & 5046.16 & 1.9 $\pm$ 0.3 \\
        \textit{Gaia} DR2   & $G$ & 6226.22 & 1.183 $\pm$ 0.015 \\
        \textit{Gaia} DR2   & $G_\mathrm{RP}$ & 7724.62 & 0.84 $\pm$ 0.09 \\[0.1cm]
        \textit{Gaia} EDR3  & $G_\mathrm{BP}$ & 5035.99 & 2.2 $\pm$ 0.3 \\
        \textit{Gaia} EDR3  & $G$ & 5822.34 & 1.468 $\pm$ 0.018 \\
        \textit{Gaia} EDR3  & $G_\mathrm{RP}$ & 7620.55 & 1.20 $\pm$ 0.11 \\[0.1cm]
        SDSS (DR16)         & $u$ & 3519.02 & 4.2 $\pm$ 0.3 \\
        SDSS (DR16)         & $g$ & 4819.97 & 2.01 $\pm$ 0.06 \\ 
        SDSS (DR16)         & $r$ & 6246.98 & 1.24 $\pm$ 0.05 \\
        SDSS (DR16)         & $i$ & 7634.91 & 0.71 $\pm$ 0.05 \\
        SDSS (DR16)         & $z$ & 9017.94 & 0.45 $\pm$ 0.12 \\[0.1cm] 
        Pan-STARRS          & $g$ & 4772.25 & 2.16 $\pm$ 0.06 \\
        Pan-STARRS          & $r$ & 6125.71 & 1.62 $\pm$ 0.05 \\
        Pan-STARRS          & $i$ & 7479.85 & 0.76 $\pm$ 0.04 \\
        Pan-STARRS          & $z$ & 8652.02 & 0.71 $\pm$ 0.06 \\
        Pan-STARRS          & $y$ & 9596.43 & 0.68 $\pm$ 0.11 \\[0.1cm]
        UKIDSS              & $Y$ & 10304.98 & 0.32 $\pm$ 0.06 \\       
        UKIDSS              & $J$ & 12501.18 & 0.43 $\pm$ 0.07 \\       
	\hline \hline
  \end{tabular}\label{tab:photometry}
\end{table}

We acquired available photometry of the object (Table~\ref{tab:photometry}) from the catalogs and surveys collected by VizieR \citep{vizier2000}. 
The data is reflected in Fig.\,\ref{fig:sed} along with the \SDSSV\ and GTC spectra of the object. The SDSS spectrum is very noisy and has, therefore, been smoothed using a three-point box car filter.  

We performed a spectral fit to the \SDSSV\ data using a model that accounts for the different light sources in the system, namely the white dwarf, the donor star, and the accretion disk. For the white dwarf, we used a grid of synthetic spectra generated using \textsc{tlusty} and \textsc{synspec} \citep{Hubeny1988,HubenyLanz1995}, covering the effective temperature range $T_\mathrm{wd} = 9000 - 40\,000\,$K in steps of 100\,K, $\log g = 8.3-8.7$ in step of 0.1\,dex (i.e. around the value $\log(g)\simeq8.5$, as expected from the mass estimate in Section~\ref{subsubsec:gr}), and a fixed metalicity, $Z = 0.5\,Z_\odot$. The white dwarf metallicity\footnote{The metallicity was estimated with a least-square fit to the \ion{Mg}{II} line with synthetic models for different metallicities: $Z = 0.01, 0.10, 0.20, 0.50, 1.00$ times their solar values. The models with $Z = 0.5\,\mathrm{Z}_\odot$ returned the lowest chi-square value.} only represents a lower limit for the metallicity of the accretion flow, which is stripped from the donor photosphere. Therefore, to model the donor, we retrieved the grid of \textsc{BT-SETTL (CFITS)} models (including a cloud model, \citealt{Allard+2003,Caffau+2011}) for late-type stars from the Theoretical Spectra Web Server\footnote{\url{http://svo2.cab.inta-csic.es/theory/newov2/index.php?models=bt-settl-cifist}}), covering $T_\mathrm{donor} = 1200 - 7000\,$K in steps of 100\,K, for $\log g_\mathrm{donor} = 3-5.5$ in steps of 0.5\,dex, for $Z = Z_\odot$.
Finally, we approximated the disk emission using an isothermal and isobaric pure-hydrogen slab model (described in \citealt{Gaensicke+1997,Gaensicke+1999}), which is defined by five free parameters: temperature, gas pressure, rotational velocity, inclination, and geometrical height.

We performed the spectral fit using the Markov chain Monte Carlo implementation in Python \texttt{emcee} \citep{Foreman-Mackey+2013}. We assumed flat priors within the grid ranges for the white dwarf temperature and surface gravity, the donor temperature and surface gravity, and the slab parameters.

The distance ($d$) to the system is an essential parameter in the spectral fitting since it provides a constraint on the white dwarf radius:
\begin{equation}\label{eq:scaling_factor}
S = \left( \frac{\Rwd}{d} \right)^2
\end{equation}
where $S$ is the scaling factor between the observed spectrum and the best-fitting model. The distance to V498\,Hya can be derived from its \textit{Gaia} parallax, which, after correction for the zero-point (see \citealt{Lindegren+2021}) results in $\varpi = 3.4 \pm 1.9\,$mas. The large ($>20\,$ percent) uncertainty of the parallax can introduce bias if the distance is computed by simple inversion \citep{Bailer-Jones2015,Luri+2018}. Therefore, we derived the distance to the system following \cite{2020MNRAS.494.3799P}, i.e., using an exponentially decreasing volume density prior and a scale height of 450\,pc, which resulted in $d=912^{+708}_{-454}$\,pc. 

Nevertheless, to account for these large uncertainties in our fitting procedure, we constrained the white dwarf and the donor to be located at the same distance $d$, which we used as a prior probability density function to convert the parallax into the distance.
For this distance range, the reddening results in $E(B-V) = 0.39$ (as derived from the 3D map of interstellar dust reddening based on PanSTARRS1 and 2MASS photometry, \citealt{Green+2019}), which we used to redden the models.

Ideally, under the assumption of a mass-radius relationship, the white dwarf mass can be constrained from Equation~\ref{eq:scaling_factor}. However, the large uncertainty of the distance implies that it is not possible to obtain reliable constraints on the white dwarf radius. To obviate this problem, in our fitting procedure, we assumed the gravitational redshift from Section~\ref{subsubsec:gr} and, allowing the effective temperature as the only free parameter for the white dwarf\footnote{During the fit, the effective temperature and the gravitational redshift constrain the $\log(g)$ of the white dwarf via the mass-radius relationship. Therefore, the $\log(g)$ of the white dwarf is allowed to vary to account for the fact that the mass-radius relationship is temperature-dependent.}, we derived its mass and radius using the mass-radius relationship from the La Plata group\footnote{\url{http://evolgroup.fcaglp.unlp.edu.ar/TRACKS/newtables.html}} \citep{Camisassa+2016}.

To account for the uncertainty of the mass ratio (see Section~\ref{subsec:q}), we included it as a free parameter in our fitting procedure and assumed a flat prior in the range $q = 0.034-0.078$. 
Knowledge of the white dwarf parameters (\Twd, \Mwd, \Rwd), the orbital period \Porb, the distance $d$, and the mass-ratio $q$ allows us to constrain the donor and slab parameters, which are computed by our fitting procedure as follows:
\begin{enumerate}
    \item the mass of the donor is given by $M_\mathrm{donor} = q\Mwd$;
    \item Kepler's third law can now be used to derive the orbital separation $a$;
    \item $q$ and $a$ yield the Roche-lobe radius, which corresponds to the radius of the donor:
    \begin{equation}\label{eq:RL}
    R_\mathrm{donor} = \frac{a 0.49 q^{2/3}}{0.6q^{2/3} + \ln(1 + q^{1/3})}
    \end{equation}
    \item $M_\mathrm{donor}$ and $R_\mathrm{donor}$ provide the surface gravity of the donor $\log g_\mathrm{donor}$;
    \item the emission from the donor is scaled to account for the irradiation from the white dwarf (equation~6 from \citealt{Pala+2019} and references therein);
    \item $q$ and $a$ also provide the circularisation radius $r_\mathrm{circ}$ and the tidal truncation radius $r_\mathrm{tidal}$ of the disc \citep{2020A&A...642A.100N}:
    \begin{equation}
        \begin{array}{l}
             r_\mathrm{circ} = a\,0.0859\,q^{-0.426}\\
             r_\mathrm{tidal} = \displaystyle a\,(0.353 + 0.271 e^{-3.045\displaystyle q})              
        \end{array}
    \end{equation}
    \item the emission of the slab is then scaled, allowing a flat prior for its radius in the range $r_\mathrm{circ} < r_\mathrm{slab} < r_\mathrm{tidal}$.
\end{enumerate}

To better constrain the fit, we also included as free parameters the fluxes of the slab emission lines of H$\alpha$, H$\beta$, and H$\gamma$, and the near-infrared observations in $Y$ and $J$ bands obtained by UKIDSS, to complement the wavelength range covered by the spectroscopic observations\footnote{The other photometric points were not included in the fit since they are in agreement with the GTC and SDSS spectroscopic data and, therefore, did not provide additional information for the fitting procedure.}.  The free parameters of our fitting procedure and their range of variation are summarised in Table~\ref{tab:fit_params} together with the best-fitting parameters.
The best-fitting model is shown in Fig.\,\ref{fig:sed} and returns $\Twd = 18\,100 \pm 150$\,K, along with $q = 0.048 \pm 0.003$ and an estimate for the distance of $d = 768 \pm 6$\,pc.

 \citet{10.1111/j.1365-2966.2010.17881.x} has shown that superoutbursts in short-period CVs are sort of a "standard candle" since most of the accretion energy is released in the superoutburst.  Equation 8 in \citet[][]{10.1111/j.1365-2966.2010.17881.x} provides the distance, assuming the measured V-magnitude corresponds to the plateau of the superoutburst. Using the data provided by T. Kato obtained around JD\,2454491 (plateau), we fetch $d=875 \pm 90$\, pc, confirming the distance by an independent method.  

Such distance is not very common for low-mass, low-accretion rates period-bouncers, but AT 2021afpi listed in 
\citep{Muñoz-Giraldo_2024} just under a distance 950 pc resembles V498\,Hya in some ways \citep{2024PASJ..tmp...86T}. By the way, the latter authors place AT 2021afpi at 720 pc, and on the borderline of period-bouncer mass ratio.

We note that the \Twd\ derived from the SED analysis should be considered as an indicative estimate, with the possibility that the white dwarf is much cooler. Assuming the best-fitting surface gravity, $\log(g)=8.47$, the wings of the Balmer absorption lines from the white dwarf could be consistent with $T_{\mathrm{wd}} \simeq 13\,000$\,K, as observed in other period bounce systems. However, the value we derived from the optical fit depends on the modeling of the accretion disc. While the slab represents a plausible approximation for the overall disk emission, it does not account for the temperature gradient in the disc. Additionally, our model does not include a model for the bright spot, which is detected as a significant contributor to emission lines (Section~\ref{sec:period}) and is also expected to contribute to the optical emission of the system.

\begin{figure}
\includegraphics[width=\columnwidth]{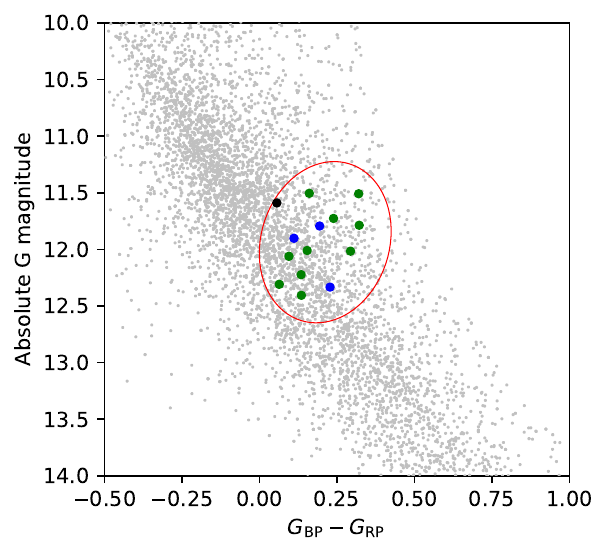}
    \caption{Fig.\,8 from \citet{2023MNRAS.525.3597I} additionally showing V498\,Hya (black dot) based on synthetic magnitudes derived from the best fit and an assumed distance of $773$\,pc. This HR diagram plots period bouncers (green dots) and candidates (blue dots) with reliable parallaxes ($\Delta \varpi < 0.2 \times \varpi$) taken from \citet{2023MNRAS.524.4867I} and \citet{2023MNRAS.525.3597I} (see Table\,4 in \citealt{2023MNRAS.525.3597I}). The red contour shows the minimal enclosing ellipse containing the known period bouncers, allowing for $1\,\sigma$ uncertainties in their \textit{Gaia} parameters. The grey dots are  the targets of the \texttt{mwm\_wd} carton within the footprint of the 236 \protect\SDSSV\ plates analysed in \citet{2023MNRAS.525.3597I}. }
    \label{fig:HR_diag_of_bouncers}
\end{figure}

In passing, we calculated synthetic \textit{Gaia} magnitudes from our best-fit model and used them to position V498\,Hya in Fig.\,8 from \citet{2023MNRAS.525.3597I}. The result (Fig.\,\ref{fig:HR_diag_of_bouncers}) shows that  V498\,Hya shares the same parameter space as other well-studied period bouncers whilst being bluer than average, which is consistent with the relatively hot white dwarf (18\,000\,K).

\begin{figure}
	\includegraphics[width=\columnwidth]{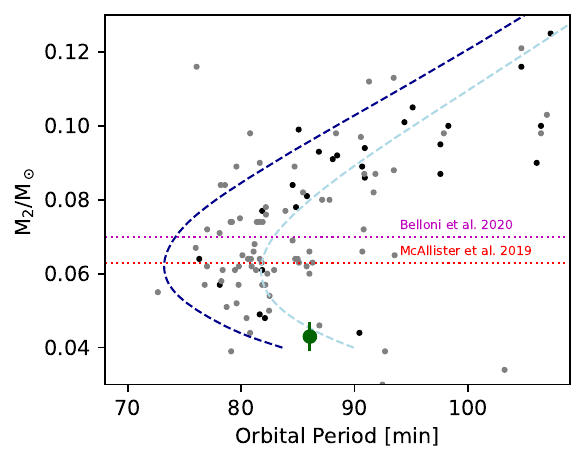}
    \caption{Donor masses in short-period CVs versus orbital periods.  The green dot with error bars reflects preferred (best-fit) solutions with \mbox{$P_{\mathrm {orb}}=86.053$\,min} and $M_\mathrm{donor}=0.043\pm0.004$\,\Msun. The black and grey points represent masses determined by eclipse modeling from \citet{2019MNRAS.486.5535M} and the stage A SH method from \citet{2022arXiv220102945K}, for which error bars have been omitted for clarity. The dashed curves represent the standard (dark blue) and optimal (light blue) evolutionary tracks in \citet{2011ApJS..194...28K}, respectively. The two horizontal lines indicate the donor mass at which the period bounce occurs, according to \citet{2019MNRAS.486.5535M} and \citet{2020MNRAS.491.5717B}.}
    \label{fig:tracks}
\end{figure}

\begin{table*}
	\centering
	\caption{Free and fixed parameters and their range of variations for the model used to fit the SED of V498\,Hya. All uncertainties are statistical.} \label{tab:fit_params}
	\begin{tabular}{lccr} 
		\hline \hline
		System parameter                   & Range covered by the models/fixed value & Best-fitting value & Constraint   \\
            \hline
            $P_\mathrm{orb}$                   & 0.05965         & - & Fixed \\
            $d$ (pc)                           & 0 -- 5000         & $768 \pm 6$ & Probability density function \\
            $v_\mathrm{grav} $ (km s$^{-1}$)   & 62.5             & - & Fixed from Section~\ref{subsubsec:gr}    \\ 
            $T_\mathrm{wd}   $ (K)             & 9000 -- 40\,000 & $18\,100 \pm 110$ &	Free flat prior \\ 
            $\Mwd   $ (\Msun)     & 0.187 - 1.35     & $0.89 \pm 0.07$   & From $v_\mathrm{grav}$ assuming a mass-radius relationship \\ 
            $\Rwd   $ (\Rsun)     & 0.0226 - 0.0032  & $0.0091 \pm 0.0009$ & From $v_\mathrm{grav}$ assuming a mass-radius relationship \\
            $\log g_\mathrm{WD}$               & 8.3 -- 8.7            & $8.47 \pm 0.11$   & From \Mwd\ and \Rwd \\
            $q$                                & 0.04 -- 0.073     & $0.048 \pm 0.003$ &	Free flat prior within the range defined in Section~\ref{subsec:q}    \\
            $T_\mathrm{donor}$ (K)             & 1200 -- 7000      & $1818 \pm 250$    &	Free flat prior \\   
            $M_\mathrm{donor}$ (M$_\odot$)     & 0.03 -- 1.2       & $0.043 \pm 0.004$ & $qM_\mathrm{WD}$ \\ 
            $R_\mathrm{donor}$ (R$_\odot$)     & 0.118 -- 0.154    & $0.105 \pm 0.004$ & Fixed to the Roche-lobe radius (Eq.~\ref{eq:RL})\\
            $\log g_\mathrm{donor}$            & 3 -- 5.5          & $5.03 \pm 0.03$   & From $M_\mathrm{donor}$ and $R_\mathrm{donor}$ \\
            $r_\mathrm{slab}$ (R$_\odot$)      & 0.2 -- 0.46       & $0.36 \pm 0.01$   & $r_\mathrm{circ} < r_\mathrm{slab} < r_\mathrm{tidal}$\\            
            $T_\mathrm{slab}$ (K)              & 5700 -- 8000      & $6400 \pm 100 $    & Free flat prior \\ 
            Slab pressure (dyn cm$^{-2}$)      & 0 -- 1000         & $140 \pm 15$      & Free flat prior \\ 
            Slab rotational velocity (km s$^{-1}$)  & 0 -- 3000         & $1000 \pm 50$       & Free flat prior \\
            Slab geometrical height (cm)       & $10^6 - 10^{12}$ & $(2.0 \pm 0.2)\times 10^8$ & Free flat prior \\
		\hline \hline
	\end{tabular}
\end{table*}

\section{Conclusions}

We have used spectroscopic time-resolved observations and survey photometry of the CV V498\,Hya to estimate the system parameters, particularly the donor mass.  

We explored multiple combinations of the SH and orbital periods to evaluate the mass ratio 
$q$. Based on the preferred SH period and S-wave-derived orbital period (P$_{\mathrm{orb}}=86.05 $\,min), we estimate $q=0.048\pm0.003$. The inferred donor mass (M$_{\mathrm{donor}}=0.043\pm0.004$\,Msun) supports the classification of V498 Hya as a period bouncer.

However, given the data's limitations, we acknowledge the tentative nature of these results. The alternative SH period yields a higher mass ratio ($ 0.074$\,\Msun), but remains consistent with a low-mass donor. While these uncertainties impact the precision of our evolutionary interpretation, the overall conclusion that V498\,Hya has evolved beyond the period minimum remains robust.
Worth noting that the deduced  \Mwd = $0.89 \pm 0.07$ \Msun\ appears higher than the average white dwarf mass in CVs \citep{2022MNRAS.510.6110P}.

\citet{2017gacv.workE..34N} has demonstrated that pre-bounce CVs usually exhibit near-IR excess around $1\mu$, and we detect no such excess in V498\,Hya. Our analysis, combining the superhumps period and the radial velocities, provides strong constraints on the white dwarf mass and mass ratio of the binary, resulting in a $0.043\pm0.004$\,\Msun\ brown dwarf as a donor star in this binary. That firmly places V498\,Hya among not-numerous period bouncers marked as a green dot in Fig.\,\ref{fig:tracks}.

Additional support is provided by the fact that this object lies inside the ellipse defined by \citet{2023MNRAS.525.3597I}  as the space containing period bouncer systems, see Fig.\,\ref{fig:HR_diag_of_bouncers}.  Revealing an additional period-bouncer among the sample discussed by \citet{2023MNRAS.525.3597I} is an important result. However, it does not change their global deficiency or affect the conclusion drawn in that paper.

The uncertainties in key parameters, such as the systemic velocity and gravitational redshift, directly influence the derived mass ratio and donor mass. Despite these challenges, our conservative approach to error propagation ensures that the primary conclusion -- V498\,Hya has evolved beyond the period minimum -- remains robust. Further high-resolution observations will be necessary to refine these estimates and confirm the system’s status as a period bouncer.

\section*{Acknowledgements}
We are grateful to Vitaly Neustroev, who reviewed this paper, for constructive input and valuable suggestions.
This project has received funding from the European Research Council (ERC) under the European Union’s Horizon 2020 Framework Programme (grant agreement no. 101020057).
This research was supported in part by the National Science Foundation under Grant No. NSF PHY-1748958 to the Kavli Institute for Theoretical Physics (KITP).
G.T. was supported by grants IN109723 and IN110619 from the Programa de Apoyo a Proyectos de Investigación e Innovación Tecnológica (PAPIIT).
This research has made use of the VizieR catalogue access tool, CDS, Strasbourg, France \citep{10.26093/cds/vizier}. The original description of the VizieR service was published in \citet{vizier2000}.
This work is (partly) based on data obtained with the instrument OSIRIS, built by a Consortium led by the Instituto de Astrofísica de Canarias in collaboration with the Instituto de Astronomía of the Universidad Autónoma de México. OSIRIS was funded by GRANTECAN and the National Plan of Astronomy and Astrophysics of the Spanish Government.
We also used publicly available data from the Catalina Sky Survey, funded by the National Aeronautics and Space Administration under Grant No. NNG05GF22G issued through the Science Mission Directorate Near-Earth Objects Observations Program. The CRTS survey is supported by the U.S. National Science Foundation under grants AST-0909182 and AST-1313422.
The UHS is a partnership between the UK STFC, the University of Hawaii, the University of Arizona, Lockheed Martin, and NASA.
\textsc{IRAF}, the Image Reduction and Analysis Facility, is distributed by the National Optical Astronomy Observatory, which is operated by the Association of Universities for Research in Astronomy under a cooperative agreement with the National Science Foundation.

\section*{Data Availability}

SDSS-V data will be publicly available at the end of the proprietary period. The GTC spectra presented here will be shared upon reasonable request to the corresponding author. The other data used in this article are available from the sources referenced in the text.



\bibliographystyle{mnras}
\bibliography{v498} 
\end{document}